\documentclass[fleqn,10pt]{wlscirep}
\usepackage[utf8]{inputenc}
\usepackage[T1]{fontenc}
\usepackage{multirow}
\usepackage{hyperref}

\title{A test case for application of convolutional neural networks to spatio-temporal climate data: Re-identifying clustered weather patterns}

\author[1]{Ashesh Chattopadhyay}
\author[1,*]{Pedram Hassanzadeh}
\author[1,2]{Saba Pasha}

\affil[1]{Rice University, Houston, 77005, USA}
\affil[2]{University of Pennsylvania, Philadelphia, 19104, USA}

\affil[*]{pedram@rice.edu}

\begin{abstract}
Convolutional neural networks (CNNs) can potentially provide powerful tools for classifying and identifying patterns in climate and environmental data. However, because of the inherent complexities of such data, which are often spatio-temporal, chaotic, and non-stationary, the CNN algorithms must be designed/evaluated for each specific dataset and application. Yet to start, CNN, a supervised technique, requires a large labeled dataset. Labeling demands (human) expert time, which combined with the limited number of relevant examples in this area, can discourage using CNNs for new problems. To address these challenges, here we 1) Propose an effective auto-labeling strategy based on using an unsupervised clustering algorithm and evaluating the performance of CNNs in re-identifying these clusters; 2) Use this approach to label thousands of daily large-scale weather patterns over North America in the outputs of a fully-coupled climate model and show the capabilities of CNNs in re-identifying the 4 clustered regimes. The deep CNN trained with $1000$ samples or more per cluster has an accuracy of $90\%$ or better. Accuracy scales monotonically but nonlinearly with the size of the training set, e.g. reaching $94\%$ with $3000$ training samples per cluster. Effects of architecture and hyperparameters on the performance of CNNs are examined and discussed.
\end{abstract}

\begin{document}

\flushbottom
\maketitle
%
%
\thispagestyle{empty}

\section*{Introduction}
Classifying and identifying specific patterns in the spatio-temporal climate and environmental data are of great interest for various purposes such as finding circulation regimes and teleconnection patterns \cite{mo1988cluster,thompson1998arctic,smyth1999multiple,bao2015cluster,sheshadri2017propagating}, identifying extreme-causing weather patterns \cite{grotjahn2016north,barnes2012methodology,vigaud2018multiscale}, studying the effects of climate change \cite{corti1999signature,barnes2014exploring,horton2015contribution,hassanzadeh2015blocking}, understanding ocean-atmosphere interaction \cite{fereday2008cluster,mckinnon2016long,anderson2017persistent}, and investigating air pollution transport \cite{zhang2012impact,souri201615}, just to name a few. Such classifications are often performed by employing empirical orthogonal function (EOF) analysis, clustering algorithms (e.g., K-means, hierarchical, self-organizing maps), linear regression, or specifically designed indices, such as those used to identify atmospheric blocking events. Each approach suffers from some major shortcomings (see the reviews by Grotjahn \textit{et al.}\cite{grotjahn2016north} and Monahan \textit{et al.}\cite{monahan2009empirical}); for example, there are dozens of blocking indices which frequently disagree and produce conflicting statistics on how these high-impact extreme-causing weather patterns will change with climate change \cite{barnes2014exploring,woollings2018blocking}.

In recent years, applications of machine learning methods for accelerating and facilitating scientific discovery have increased rapidly in various research areas. For example, in climate science, neural networks have produced promising results for parameterization of convection and simulation of clouds \cite{schneider2017earth,gentine2018could,brenowitz2018prognostic,rasp2018deep,o2018using}, weather forecasting \cite{rasp2018neural,dueben2018challenges}, and predicting {El N}i\~no \cite{Nooteboom2018}. A class of supervised deep learning architectures, called convolutional neural network (CNN), has transformed pattern recognition and image processing in various domains of business and science\cite{lecun2015deep,goodfellow2016deep} and can potentially become a powerful tool for classifying and identifying patterns in the climate and environmental data. In fact, in their pioneering work, Liu \textit{et al.} \cite{liu2016application} and Racah \textit{et al.} \cite{racah2016semi} have shown the promising capabilities of CNNs in identifying tropical cyclones, weather fronts, and atmospheric rivers in large, labeled climate datasets. 

Despite the success in applying CNNs in these few studies, to further expand the applications and usefulness of CNNs in climate and environmental sciences, there are some challenges that should be addressed\cite{karpatne2018machine}. One major challenge is that unlike the data traditionally used to develop and assess CNN algorithms such as the static images in ImageNet \cite{krizhevsky2012imagenet}, the climate and environmental data, from model simulations or observations, are often spatio-temporal and highly nonlinear, chaotic, high-dimensional, non-stationary, multi-scale, and correlated. For example, the large-scale atmospheric circulation, whose variability strongly affects day-to-day weather and extreme events, is a high-dimensional turbulent system with length scales of $100$~km to $10000$~km and time scales of minutes to decades (and beyond), with strongly coherent and correlated patterns due to various physical processes, and non-stationarity due to, e.g., atmosphere-ocean coupling and anthropogenic effects \cite{williams2017census,ma2017quantifying,kosaka2013recent}. An additional challenge with observational datasets is that they are usually short and sparse and have measurement noise.

As a result, to fully harness the power of CNNs, the algorithms (architecture, hyperparameters etc.) have to be designed and evaluated for each specific climate or environmental data and for each specific application. However, to start, CNN, as a supervised technique, requires a large labeled dataset for training/testing. Labeling data demands (human) expert time and resources and while some labeled datasets for specific types of data and applications are now publicly available \cite{racah2016semi,prabhat2015teca}, can discourage exploring the capabilities of CNN for various problems. With these challenges in mind, the purpose of this paper is twofold:
\begin{enumerate}
\item To propose an effective, simple approach for labeling any spatio-temporal climate and environmental data based on using K-means clustering, which is an easy-to-implement, unsupervised classification technique,   
\item To use this approach and label thousands of large-scale weather patterns over North America in the outputs of a state-of-the-art climate model and show the capabilities of CNNs in identifying the four different clusters and examine how the performance of CNNs depend on the architecture, hyperparameters, and size of the training dataset. 
\end{enumerate}

\section*{Methodology}
The approach proposed here involves two steps: i) the spatio-temporal data is clustered into $n$ classes using an unsupervised technique such as K-means\cite{lloyd1982least}, which assigns an index ($1$ to $n$) to each pattern in the dataset, and ii) the cluster indices are used to label the patterns in the dataset, $1$ to $n$. The labeled dataset is then used to train and test the CNN. The performance of CNN in re-identifying the cluster indices of patterns in the testing phase can be used to evaluate and improve the CNN algorithms for each specific dataset. Note that while here we use K-means clustering for indexing, other algorithms such as hierarchical, expectation-maximization, or self-organizing maps\cite{cheng1993cluster,smyth1999multiple,bao2015cluster} can be used instead. However, the K-means algorithm, which clusters the data into {\it a priori} specified $n$ based on Euclidean distances, provides an effective, simple method for the objective here, which is to label the dataset for evaluating CNN, rather than finding the most meaningful (if possible\cite{fereday2008cluster}) number of clusters in the spatio-temporal data.    

The approach proposed here can be used for any climate or environmental data such as wind, precipitation, or sea-surface temperature patterns or distributions of polluters, to name a few. For the case study presented here, we focus on the daily weather patterns over North America in summer and winter. The data and K-means clustering and CNN algorithms are presented in Data and Methods, but their key aspects are briefly discussed below. We use data from the Large Ensemble (LENS) Community Project \cite{kay2015community}, which consists of a $40$-member ensemble of fully-coupled Community Earth System Model version 1 (CESM1) simulations with historical radiative forcing from $1920$ to $2005$. We focus on daily averaged geopotential height at $500$~hPa (Z500 hereafter), whose isolines are approximately the streamlines of the large-scale circulation at mid-troposphere and are often used to represent weather patterns \cite{holton2012introduction}. Daily Z500 from $1980$ to $2005$ provides $\sim 95000$ samples for summer months and for winter months over North America.

 \begin{figure}[t]
  \centering
  \includegraphics[width=.9\textwidth]{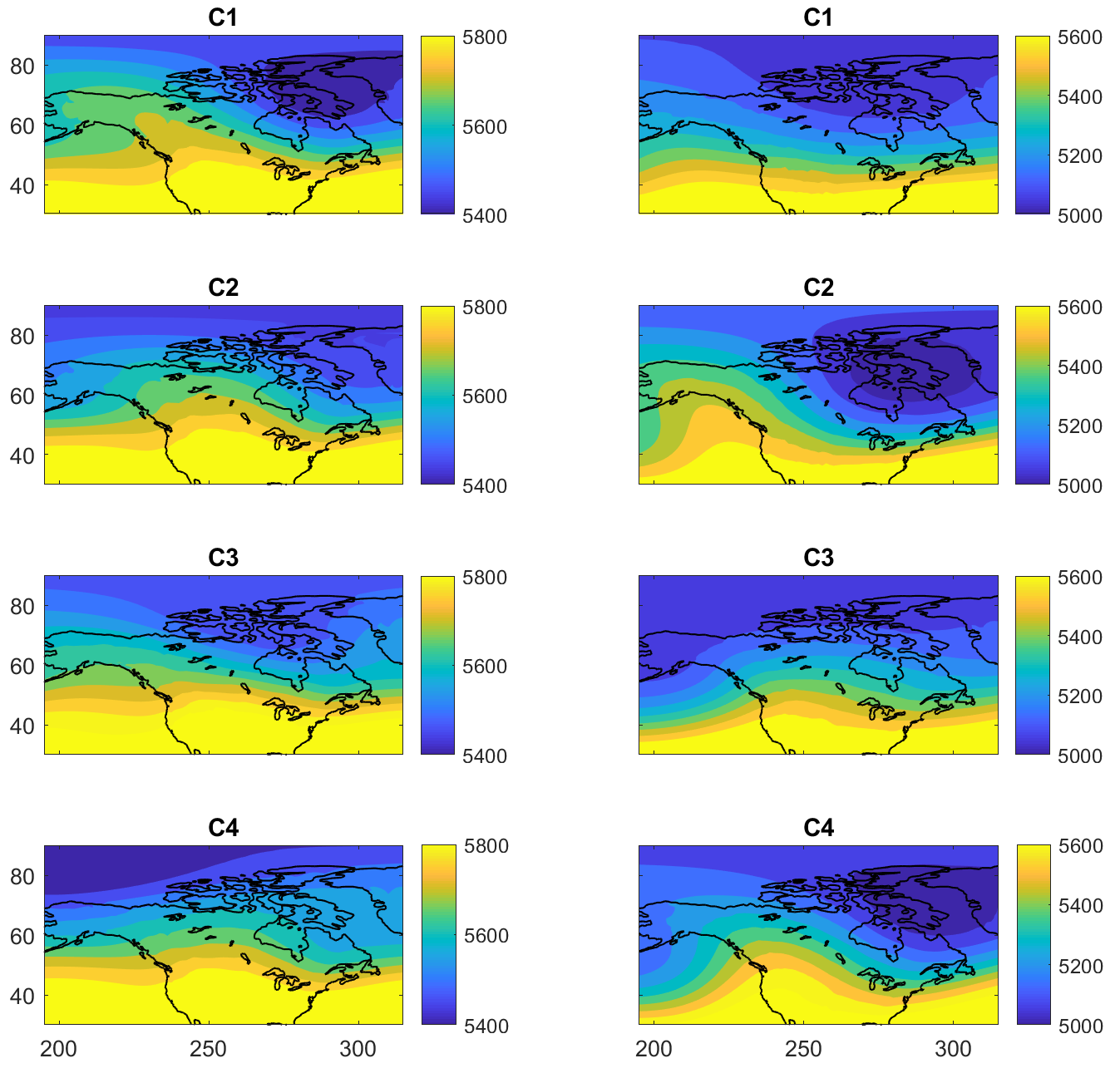}
  \caption{Centers of the four K-means clusters in terms of the full Z500 field (with unit of meters) for summer months, June-August (left column) and for winter months, December-February (right column). The K-means algorithm finds the cluster centers based on {\it a priori} specified number of clusters $n$ ($=4$ here) and assigns each daily pattern to the closest cluster center based on Euclidean distances. The assigned cluster indices are used as labels for training/testing CNNs. Note that K-means clustering is performed on daily zonal-mean-removed Z500 anomalies projected onto their first 22 EOFs, but the cluster indices are used to label the full Z500 patterns to minimize pre-processing and retain the complex temporal variabilities of the Z500 field (see Data and Methods for further discussions).}
  \label{fig:Zcenters}
\end{figure}

As discussed in Data and Methods, the K-means algorithm is used to classify the winter days and summer days (separately) into $n=4$ clusters. The clustering analysis is performed on zonal-mean-removed daily Z500 anomalies projected on 22 EOFS that retain approximately $95\%$ of the variance; however, the computed cluster index for each day is used to label that day's full Z500 pattern. The four cluster centers in  terms of the full Z500 field are shown in Figure \ref{fig:Zcenters}. Labeled full Z500 patterns are used as input to CNN for training and testing. We work with the full Z500 fields, rather than the computed anomalies, because one hopes to use CNN with minimally pre-processed data. Therefore, we focus on the more difficult task of re-identifying the clusters in the full Z500 fields, which include the complex temporal variabilities such as the seasonal cycle and non-stationarity resulting from the low-frequency coupled ocean-atmosphere modes and changes in the radiative forcing between $1980$ and $2005$. We further emphasize that the spatio-temporal evolution of Z500 field is governed by high-dimensional, strongly nonlinear, chaotic, multi-scale dynamics\cite{holton2012introduction}.   

\begin{figure}[t]
  \centering
    \includegraphics[width=1.0\textwidth]{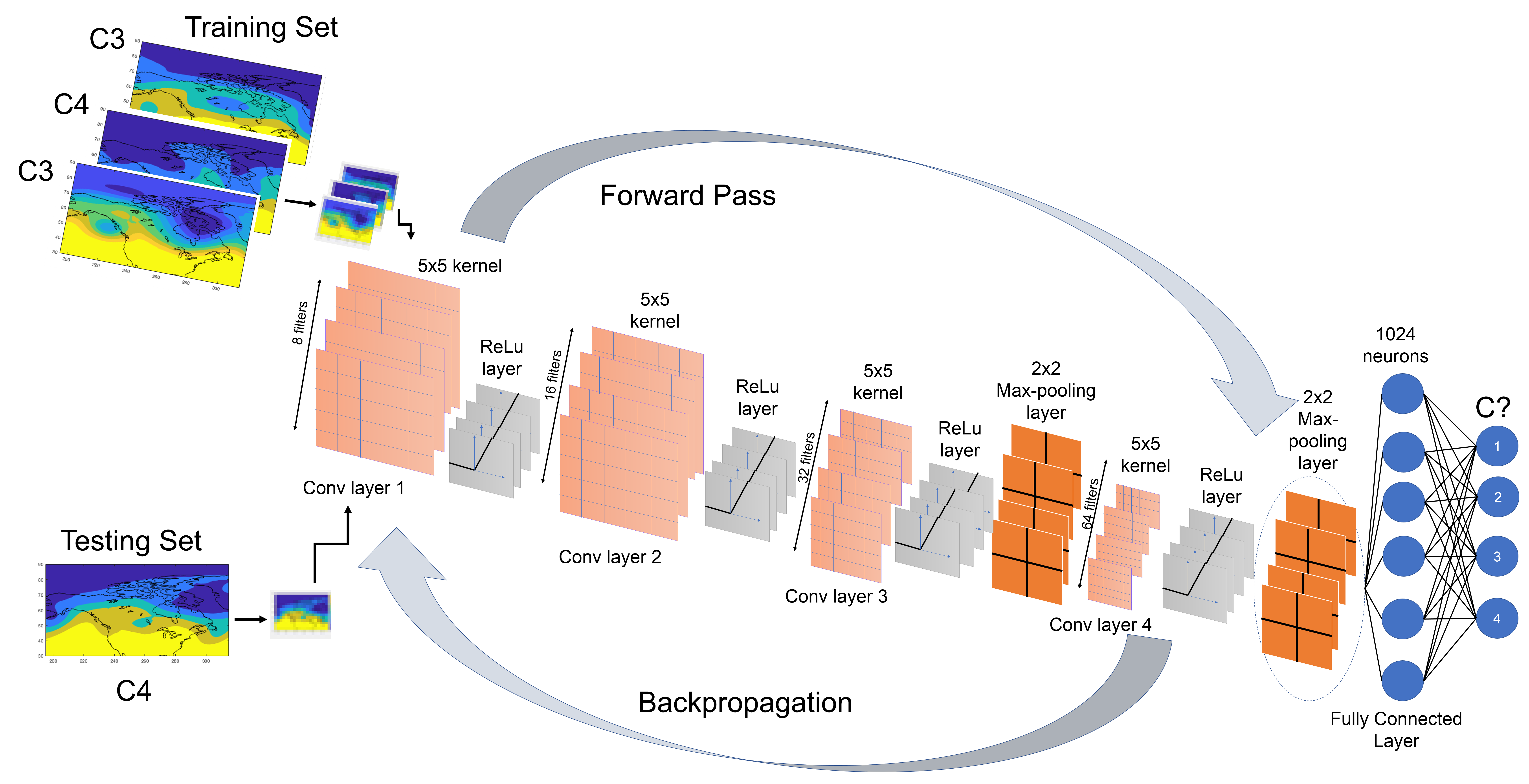}
      \caption{The architecture of CNN4, which has $4$ convolutional layers that have $8,16,32$ and $64$ filters, respectively. Each filter has a kernel size of $5 \times 5$. Filters of the max-pooling layer have a kernel size of $2 \times 2$. The convolutional layers at the beginning capture the low-level features  while the latter layers would pick up the high level features \cite{girshick2014rich}. Each convolution step is followed by the ReLU layer that introduces nonlinearity in the extracted features. In the last two layers, a max-pooling layer after the ReLU layer  retains only the most dominant features in the extracted feature map while inducing translation and scale invariance. These extracted feature maps are then concatenated into a single vector which is connected to a fully connected neural network with $1024$ neuron. The output is the probability of each class. The input images into this network have been first down-sampled using bi-cubic interpolation to only retain the large-scale features in the circulation patterns (Figure~\ref{fig:sampling}). }
      \label{fig:arch}
\end{figure}

\begin{figure}[t]
  \centering
    \includegraphics[width=1.0\textwidth]{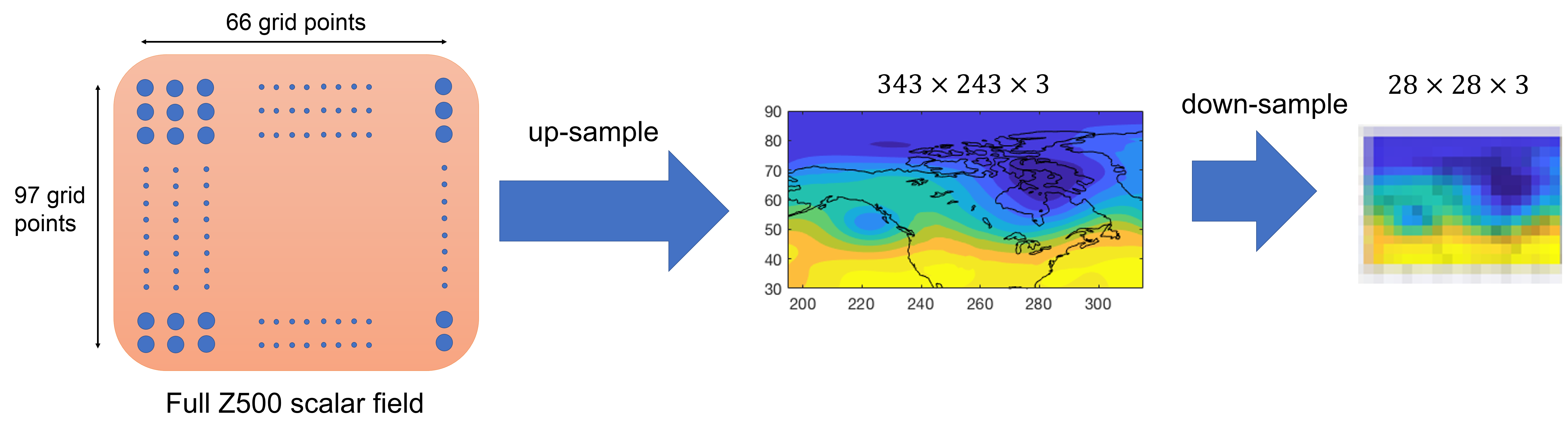}
      \caption{Schematic of the up-sampling and down-sampling steps. Each daily full Z500 pattern, which is on a $66 \times 97$ latitude-longitude grid, is converted to a contour plot represented by a RGB image of size $ 342 \times 243$ pixels with $3$ channels representing red, green, and blue. This up-sampled image is then down-sampled to an image of size  $28 \times 28 \times 3$ using bi-cubic interpolation and further standardized by subtracting the mean and dividing by the standard deviation of the pixel intensities. These images are the inputs to CNN for training or testing. The down-sampling step is used to remove redundant features at small scales from each sample. Trying to learn such small features, which are mostly random, can result in overfitting of the network (see Data and Methods for further discussions).}
      \label{fig:sampling}
\end{figure}

The architecture of our CNN algorithm is shown in Figure \ref{fig:arch}. In general, the main components of a CNN algorithm are: convolutional layers in which a specified number of kernels (filters) of specified sizes are applied to extract the key features in the data and produce feature maps;  Rectified Linear Unit (ReLU) layers in which the ReLU activation function, $f(x)=\max(0,x)$, is applied to the feature maps to introduce nonlinearity; pooling layers that reduce the dimensions of the feature maps to increase the computational performance, control overfitting, and induce translation and scale invariance (which is highly desirable for the chaotic spatio-temporal data of interest here); and finally, fully connected layers \cite{lecun2015deep,goodfellow2016deep}. The inputs to CNN are the full Z500 fields that are converted to images and down-sampled to reduce redundancies in small scales (Figure~\ref{fig:sampling}). During the training phase, the images and their cluster indices, from a randomly drawn training set (TR), are inputted into CNN and the kernels (i.e. their weights) are learned using backpropagation \cite{goodfellow2016deep}. {The major advantage of CNNs over traditional image processing methods is that the appropriate kernels are learned for each dataset, rather than being hand-engineered and specified {\it a priori}}. During the testing phase, images, from a randomly drawn testing set (TS) that has no overlap with TR, are inputed into the CNN and the output is the predicted cluster index. If the CNN has learned the key features of these high-dimensional, nonlinear, chaotic, non-stationary patterns, then the predicted cluster indices should be largely correct. 

In this paper we developed two CNNs, one with two convolutional layers (CNN2) and another one with four convolutional layers (CNN4). The effects of hyperparameters and other practical issues as well as scaling of the accuracy with the size of the training set are examined and discussed.

\section*{Results}
\label{results}

 \subsection*{Performance of CNN}
Tables~\ref{table:summer} and \ref{table:winter} show the test accuracies of CNN2 and CNN4 for the summer and winter months, respectively. The CNN4 has the accuracy of $93.3\% \pm 0.2\%$ (summer) and $93.8\% \pm 0.1\%$ (winter) while CNN2 has the accuracy of $89.0\% \pm 0.3\%$ (summer) and $86.6\% \pm 0.3\%$ (winter). The reported accuracies are the mean and standard deviation of the accuracies of the $5$ sets in the TS. The $4\%-7\%$ higher accuracy of the deeper net, CNN4, comes at the price of higher computational demands (time and memory) because of the two additional convolutional layers, however, the robust test accuracy of $\sim 93\%$ is significant for the complex patterns studied here. 

Deep CNNs are vulnerable to overfitting as the large number of parameters can lead to perfect accuracy on the training samples while the trained CNN fails to generalize and accurately classify the new unseen samples because the CNN has memorized, rather than learned, the classes. In order to ensure that the reported high accuracies of CNNs here are not due to overfitting, during the training phase, a randomly chosen validation set (which does not have any overlap with TS or TR) was used to tune the hyperparameters (see Data and Methods). For each case in Tables~\ref{table:summer} and \ref{table:winter}, the reported test accuracy is approximately equal to the training accuracy after the network converges, which along with small standard deviations among the $5$ independent sets in the TS, indicate that the classes have been learned rather than overfitted. It should be mentioned that for this data with the TR of size $N \le 12000$, we have found that overfitting occurs if more than $4$ convolutional layers are used.

\begin{table}[t]
\caption{The confusion matrix for CNN4 (CNN2) applied to summer months. A TR of length $N=12000$ ($3000$ samples per cluster) and a TS, consisting of $5$ independent sets each with $1000$ samples per cluster, are used (see Data and Methods). The TR and TS are randomly selected and have no overlap. Each number shows how many patterns from a given cluster in TS are identified by the trained CNN to belong to each cluster (the mean and standard deviation from the $5$ sets of TS are reported). The results are from the best trained CNN2 and CNN4. The overall test accuracy, calculated as the sum of the diagonal numbers, i.e. all correctly identified patterns, divided by the total number of patterns, i.e. $3000$, and turned to percentage is  $93.3\% \pm{0.2}\%$ (CCN4) and $89.0\% \pm{0.3}\%$ (CNN2).} 
\centering 
\begin{tabular}{c | c c c c} 
\hline\hline 
 & \textbf{Identified as C1} &\textbf{ Identified as C2} &\textbf{ Identified as C3} & \textbf{Identified as C4} \\ 
\hline 
\textbf{True C1} & $ \mathbf{915 \pm{3}} \, \left( \mathbf{959 \pm{8}}\right) $ & $30 \pm{3} \, \left(14 \pm{4}\right) $&$55 \pm{3} \, \left(27 \pm{2}\right)$ & $0 \pm{0} \, \left(0 \pm{0}\right) $\\ 
\textbf{True C2} & $ 17 \pm{2}  \, \left(78 \pm{3}\right) $ & $ \mathbf{906 \pm{3}} \, \left(\mathbf{819 \pm{2}}\right) $ & $48 \pm{2} \, \left(58 \pm{2}\right)$ & $ 29 \pm{3}\left(45 \pm{4}\right) $\\
\textbf{True C3} & $ 17 \pm{1} \, \left(61 \pm{1}\right) $ & $8 \pm{1} \, \left(30 \pm{1}\right) $ & $ \mathbf{955 \pm{3}} \, \left(\mathbf{857 \pm{3}}\right)$ & $ 20 \pm{2} \, \left(52 \pm{2}\right) $\\
\textbf{True C4} & $ 0 \pm{0} \, \left(0 \pm{0}\right) $ & $ 18 \pm{2} \, \left(49 \pm{2}\right)$ & $ 26\pm{3} \, \left(23 \pm{2}\right)$ & $ \mathbf{956 \pm{3}} \, \left(\mathbf{928\pm{3}}\right) $\\
\hline \hline  
\end{tabular}
\label{table:summer} 
\end{table}
 
\begin{table}[t]
\caption{Same as Table~\ref{table:summer} but for winter months. The overall test accuracy is $93.8\% \pm{0.1}\%$ (CNN4) and $86.6\% \pm{0.3}\%$ (CNN2).} 
\centering 
\begin{tabular}{c | c c c c} 
\hline\hline 
 & \textbf{Identified as C1} &\textbf{ Identified as C2} &\textbf{ Identified as C3} & \textbf{Identified as C4} \\ 
\hline 
\textbf{True C1} &$ \mathbf{937\pm{2}} \, \left( \mathbf{783\pm{3}}\right)$ &13 $ \, \left(98 \pm{1}\right ) $&$37\pm{1} \, \left(20 \pm{1}\right)$ &1$3 \pm{1} \, \left(99 \pm{1}\right) $\\ 
\textbf{True C2} &$ 71 \pm{2} \, \left(28 \pm{2}\right) $&$ \mathbf{920 \pm{2}} \, \left(\mathbf{951 \pm{2}}\right) $& $0 \pm{0} \, \left(0 \pm{0}\right)$ &$ 9 \pm{1} \, \left(21 \pm{1}\right) $\\
\textbf{True C3} & $ 49\pm{2} \, \left(61\pm{1}\right) $& $0 \pm{0} \, \left(0 \pm{0}\right) $& $ \mathbf{984 \pm{3}} \, \left(\mathbf{822 \pm{2}}\right)$ & $ 3 \pm{0} \, \left(129 \pm{0}\right) $\\
\textbf{True C4} &$ 23 \pm{0} \, \left(4 \pm{0}\right)$ &$ 29 \pm{2} \, \left(82 \pm{2}\right)$ & $ 37 \pm{2} \, \left(3 \pm{1}\right)$ & $ \mathbf{911 \pm{2}} \, \left(\mathbf{911 \pm{3}}\right) $\\
\hline \hline 
\end{tabular}
\label{table:winter} 
\end{table} 

\subsection*{Scaling of the test accuracy with the size of the training set}
\begin{figure}[t]
  \centering
  \includegraphics[width=0.9\textwidth]{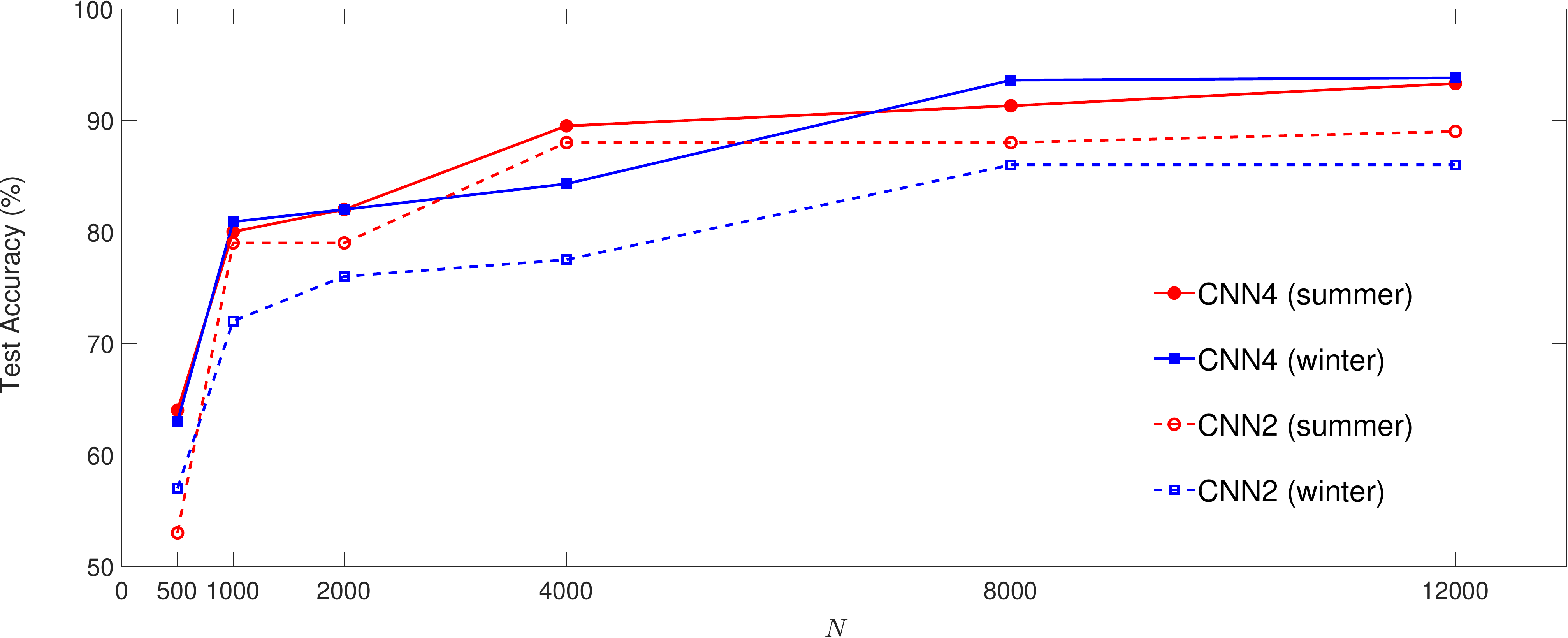}
  \caption{Test accuracy of CNN4 and CNN2 as a function of the size of the training set $N$. To avoid class imbalance, $N/4$ samples per cluster are used. A $3:1$ ratio between the number of samples per cluster in the training and testing sets are maintained.  
}
  \label{fig:scaling}
\end{figure}

An important practical question that many ask before investing in labeling data and developing their CNN algorithm is ``how much data do I need to get reasonable accuracy with CNN?''. However, a theoretical understanding of the bound or scaling of CNNs' accuracy based on the number of the training samples or number of tunable parameters of the network is currently unavailable \cite{zhang2016understanding}. Given the abundance of the labeled samples in our dataset, it is an interesting experiment to examine how the test accuracy of CNNs scales with the size of the TR, $N$. Figure~\ref{fig:scaling} shows that the test accuracy of CNN2 and CNN4 scales monotonically but nonlinearly with $N$ for summer and winter months. With $N=500$ ($125$ training samples per cluster), the test accuracy of CNN4 is around $64\%$. The accuracy jumps above $80\%$ with $N=1000$ and then increases to above $90\%$ as $N$ is increased to $8000$. Further increasing $N$ to $12000$ will slightly increase the accuracy to $93\%$. The accuracy of CNN2 qualitatively shows the same behavior, although it is consistently lower than the accuracy of CNN4 for the same $N$. While the empirical scaling presented here is very likely problem dependent and cannot replace a theoretical scaling, it provides an example of how the test accuracy might depend on the size of the TR.                    


 \subsection*{Incorrectly classified patterns } 
While the results presented above show an outstanding performance by CNN and test accuracy of $\sim 93\%$ with $N=12000$ for CNN4, the cluster indices of a few hundred testing samples (out of the $4000$) have been incorrectly identified. From visually comparing examples of correctly and incorrectly identified patterns, inspecting the cluster centers in Figure~\ref{fig:Zcenters}, or examining the results of Tables~\ref{table:summer} and \ref{table:winter}, it is not easy to understand why patterns from some clusters have been more (or less) frequently mis-classified. For example, in summer months using CNN4, patterns in cluster C2 (C4) are the most (least) frequently mis-classified. Patterns in C2 are most frequently mis-identified to belong to C3 (48 samples) while patterns in C3 are rarely mis-identified to belong to C2 (8). There are many examples of such asymmetries in mis-classification in Tables~\ref{table:summer} and \ref{table:winter}, although there are some symmetric examples too, most notably no mis-classification between C1 and C4 in summer or C2 and C3 in winter. It should be also noted that while CNN4 has robustly better overall test accuracy compared to CNN2 for summer/winter or as $N$ changes, it may not improve the accuracy for every cluster (e.g. $915$ C2 samples are  correctly identified by CNN4 for summer months compared to $959$ by CNN2). Visual inspection of cluster centers does not provide much clues on which clusters might be harder to re-identify or mix up, e.g., patterns in C2 in winter months are frequently (71 samples) mis-classified as C1 while rarely mis-classified as C3 (0) or C4 (9 samples) even though the cluster center of C2, which has a notable ridge over the eastern Pacific ocean and a low-pressure pattern over north-eastern Canada, is (visually) distinct from the cluster center of C1 or C3 but resembles that of C4. 

While understanding {\it how} a CNN  learns or {\it why} some patterns are identified and some are mis-identified can be of great interest for many applications, particularly those involved addressing a scientific problem, answering such questions is not straightforward with the current understanding of deep learning \cite{lin2017does}. In the results presented here, there are two potential sources of inaccuracy: imperfect learning and improperly labeled patterns. The former can be a result of unoptimized choice of the hyperparameters or insufficient number of training samples. As discussed in Data and Methods, we have explored a range of hyperparameters, manually for some and using optimization algorithms such as ADAM\cite{kingma2014adam} for others. Still there might be room for further systematic optimization and improvement of the test accuracy. The results of Figure~\ref{fig:scaling} suggest that increasing $N$ would have a small effect on the test accuracy. Training CNN4 for summer with $N=18000$ increases the best test accuracy from $93.3\%$ (obtained with $N=12000$) to just $94.1\%$. These results suggest that the accuracy might be still further improved, although very slowly, by increasing $N$.         

          

Another source of inaccuracy might be related to how the patterns are labeled using the K-means cluster indices. The K-means algorithm is deterministic and assigns each pattern to one and only one cluster index. In data that have well-defined classes, the patterns in each cluster are very similar to each other (high cohesion) and dissimilar from patterns in other clusters (well separated). However, in chaotic, correlated, spatio-temporal data, such as those studied here, some patterns might have similarities to more than one cluster, however, the K-means algorithm assigns them to just one (the closest) cluster. As a result, two patterns that are very similar might end up in two different clusters and thus assigned different labels. The presence of such borderline cases in the TR can degrade the learning process for CNN and their presence in the TS can reduce the test accuracy. The silhouette value $s$ is a measure often used to quantify how a pattern is similar to its own cluster and separated from the patterns in other clusters\cite{rousseeuw1987silhouettes}. Large, positive values of $s$ indicate high cohesion and strong separation, while small and in particular negative values indicate the opposite.

To examine whether part of the inaccuracy in the testing phase is because of borderline cases, in Table~\ref{table:nonlin} we show the percentage of samples correctly classified or incorrectly classified for ranges of high and negative silhouette values. The results indicate that poorly clustered (i.e. borderline) patterns, e.g. those with $s<0$, are more frequently mis-classified compared to well-clustered patterns, e.g., those with $s>0.4$ ($11\%$ versus $4.8\%$). This analysis suggests that part of the $6.7\%$ testing error of CNN4 for summer months might be attributed to poor clustering and improper labeling (one could remove samples with low $s$ from the TR and TS, but here we chose not to in order to have a more challenging task for the CNNs).


Note that soft clustering methods (e.g. fuzzy c-means clustering \cite{bezdek1984fcm}) in which a pattern can be assigned to more than one cluster might be used to overcome the aforementioned problem if it becomes a significant source of inaccuracy. In any case, one has to ensure that the labels obtained from the unsupervised clustering technique form a learnable set for the CNN and be aware of the potential inaccuracies arising from poor labeling alone.

 \begin{table}[t]

\caption{Percentage of samples correctly classified or incorrectly classified for different ranges of silhouette values, $s$. Silhouette values, by definition, are between $-1$ and $1$ and high (low and particularly negative) values indicate high (low) cohesion and strong (weak) separation. Percentages show the fraction of patterns in a given range of silhouette values. The samples are from summer months and for CNN4.} 
\centering 
\begin{tabular}{l | c c c} 
\hline\hline 
 Silhouette value &$s<0$ &$s>0.2$ & $s>0.4$  \\ 
  \hline
 Correctly identified&$89.0\%$&$94.2\%$&$95.2\%$ \\
 \hline
Incorrectly identified&$11.0\%$&$5.8\%$&$4.8\%$ \\ \hline \hline

\end{tabular}
\label{table:nonlin} 
\end{table}

\section*{Discussion}
The unsupervised auto-labeling strategy proposed here facilitates exploring the capabilities of CNNs in studying problems in climate and environmental sciences. The method can be applied to any spatio-temporal data and allows one to examine the power and limitations of different CNN architectures and scaling of their performance with the size of the training dataset for each type of data before further investing in labeling the patterns to address specific scientific problems, e.g. to study patterns that cause heat waves or extreme precipitation.  

The analysis conducted on daily large-scale weather patterns shows the outstanding performance of CNNs in identifying patterns in chaotic, multi-scale, non-stationary, spatio-temporal data with minimal pre-processing. Building on the promising results of previous studies\cite{liu2016application,racah2016semi}, our analysis goes beyond their binary classifications and shows over $90\%$ test accuracy for $4$-cluster classification once there are at least $2000$ training samples per cluster.      

The promising capabilities of CNNs in identifying complex patterns in climate data further opens frontiers for prediction of some weather, and in particular extreme weather, events using deep learning methods. Techniques such as recurrent neural networks (RNNs) with long short-term memory (LSTM) and tensor-train RNNs have shown encouraging skills in predicting time series in chaotic systems\cite{vlachas2018data,yu2017long}. Coupling CNNs with these techniques can potentially provide powerful tools for spatio-temporal prediction; e.g., a convolutional LSTM network has been recently implemented for precipitation nowcasting \cite{xingjian2015convolutional}. 


\section*{Data and Methods}
\subsection*{Data from the Large Ensemble (LENS) Community Project}
\label{data}
We use data from the publicly available Large Ensemble (LENS) Community Project \cite{kay2015community}, which consists of a 40-member ensemble of fully-coupled atmosphere-ocean-land-ice Community Earth System Model version 1 (CESM1) simulations at the horizontal resolution of $\sim 1^\mathrm{o}$. The same historical radiative forcing from $1920$ to $2005$ is used for each member; however, small, random perturbations are added to the initial state of each member to create an ensemble. We focus on daily averaged geopotential height at $500$~hPa (Z500). Z500 isolines are approximately the streamlines of the large-scale circulation at mid-troposphere and are often used to represent weather patterns \cite{holton2012introduction}. We focus on Z500 from $1980$ to $2005$ for the summer months of June-August ($92$ days per summer) for all 40 ensemble members (total of 95680 days) over North America, $30^\mathrm{o}-90^\mathrm{o}$ north and $200^\mathrm{o}-315^\mathrm{o}$ east (resulting in $66 \times 97$ latitude--longitude grid points). Similarly, for winter we use the same $26$ years of data for the months of December, January, and February ($90$ days per winter and a total of $95508$ days).

\subsection*{Clustering of weather patterns}
\label{clustering}
The daily Z500 patterns over North America are clustered for each season into $n=4$ classes. Following Vigaud et al.\cite{vigaud2018multiscale}, first, an EOF analysis is performed on the data matrix of zonal-mean-removed Z500 anomalies and the first $22$ principal components (PCs), which explain $95\%$ of the variance, are kept for clustering analysis. The K-means algorithm\cite{lloyd1982least} is used on these $22$  PCs and repeated $1000$ times with new initial cluster centroid positions, and a cluster index $k=1,2,3$ or $4$ is assigned to each daily pattern. 


It should be noted that the number of clusters $n=4$ is not chosen as an optimal number, which might not even exist for these complex, chaotic, spatio-temporal data\cite{fereday2008cluster}. Rather, for the purpose of the analysis here, the chosen $n$ should be large enough such that the cluster centers are reasonably distinct and there are several clusters to re-identify to evaluate the CNNs in a challenging multi-class classification problem, yet, small enough such that there are enough samples per cluster for training and testing. 

\subsection*{Labeling and up/down-samplings}
Once the cluster index for each daily pattern is computed, the full Z500 daily patterns are labeled using these indices. We focus on the full Z500 fields, rather than the anomalies, for several reasons: (1) The differences between the patterns from different clusters are more subtle in the full Z500 compared to the anomalous Z500 fields; (2) The full Z500 fields contain all the complex, temporal variabilities and non-stationarity resulting from ocean-atmosphere coupling and changes in the radiative forcing while some of these variabilities might be removed by computing the anomalies; As a result of (1) and (2), re-identifying the cluster indices in the full Z500 fields provide a more challenging test for CNNs; (3) One hopes to use CNNs with no or minimal pre-processing of the data, consequently, we focus on the direct output of the climate model, i.e., full Z500 field, rather than the pre-processed anomalies. 

In our algorithm, the only pre-processing conducted on the data is the up-sampling/down-sampling shown in Figure~\ref{fig:sampling}. The down-sampling step is needed to remove the small-scale, transient features of the chaotic, multi-scale atmospheric circulation from the learning/testing process. Inspecting the cluster centers in Figure \ref{fig:Zcenters} shows that the main differences between the four clusters are in large-scale. If the small-scale features, which are associated with processes such as baroclinic instability, are not removed via down-sampling, the CNN will try to learn the distinction between these features in different classes, which is futile as these features are mostly random. 
We have found in our analysis that without the down-sampling step, we could not train the CNN using a simple random normal initialization of the kernel weights (if instead of random initialization, a selective initialization method such as Xavier\cite{he2015delving} is used, the network can be trained for the full-sized images although the test accuracy remains low due to overfitting on small-scale features.) The need for down-sampling in applications of CNNs to multi-scale patterns has been reported previously in other areas \cite{lin2017feature}. In the applications that involve the opposite case, i.e. when the small-scale features are of interest and have to be learned, techniques such as localization can be used \cite{kim2017resolution}.

Note that although Z500 is a scalar field, here we have used the three channels of RGB to represent it because we are focusing on only one variable. In the future applications, when several variables are studied together, each channel can be used to represent one scalar field, e.g. temperature and/or components of the velocity vector.

 \subsection*{Convolutional Neural Network (CNN)}
The CNN is developed using the Tensorflow library\cite{abadi2016tensorflow} following the Alex Net architecture \cite{krizhevsky2012imagenet}. We have trained and tested two CNNs: one with two convolutional layers, named {CNN2}, and a deeper one with $4$ layers, called {CNN4}. 

\subsubsection*{CNN2}
The shallow neural network has two convolutional layers with $16$ and $32$ filters, respectively. Each filter has a kernel size of $5 \times 5$. In each convolutional layer, zero padding around the borders of images is used to maintain the size before and after applying the filters. Each convolutional layer is followed with a ReLU activation function and a max-pooling layer that has a kernel size of $2 \times 2$ and stride of $1$ (stride is the number of pixels the filter shifts over in the pooling layer\cite{goodfellow2016deep}). The output feature map is $7 \times 7 \times 64$ which is fed into a fully connected neural network with $1024$ neurons. The cross entropy cost function is accompanied by a $L_2$ regularization term with a hyperparameter $\lambda$. Furthermore, to prevent overfitting, dropout regularization with hyperparameter $p$ has been used in the fully connected layer. An adaptive learning rate $\alpha$, a hyperparameter, is implemented through the ADAM optimizer \cite{kingma2014adam}. The final output is the probability of the input pattern belonging to each cluster. A softmax layer assigns the pattern to the cluster index with the highest probability. 

\subsubsection*{CNN4}
The deeper neural network, CNN4, is the same as CNN2, except that there are four convolutional layers, which have $8,16,32$ and $64$ filters, respectively (Figure~\ref{fig:arch}). Only the last two convolutional layers are followed by max-pooling layers.

\subsubsection*{Training, validating, and testing procedures}
\label{training}
For the case with $N=12000$, $3000$ labeled images from each of the four clusters is selected randomly (the TR set). Separately, $4$ validation datasets, each with $1000$ samples per cluster are randomly selected. For the testing set (TS), $5$ datasets, each with $1000$ samples per cluster, are randomly selected. The TR, validation sets, and TS have no overlap. The equal number of samples from each cluster prevents class imbalance in training and testing. 

\begin{figure}[t]
  \centering
  \includegraphics[width=\textwidth]{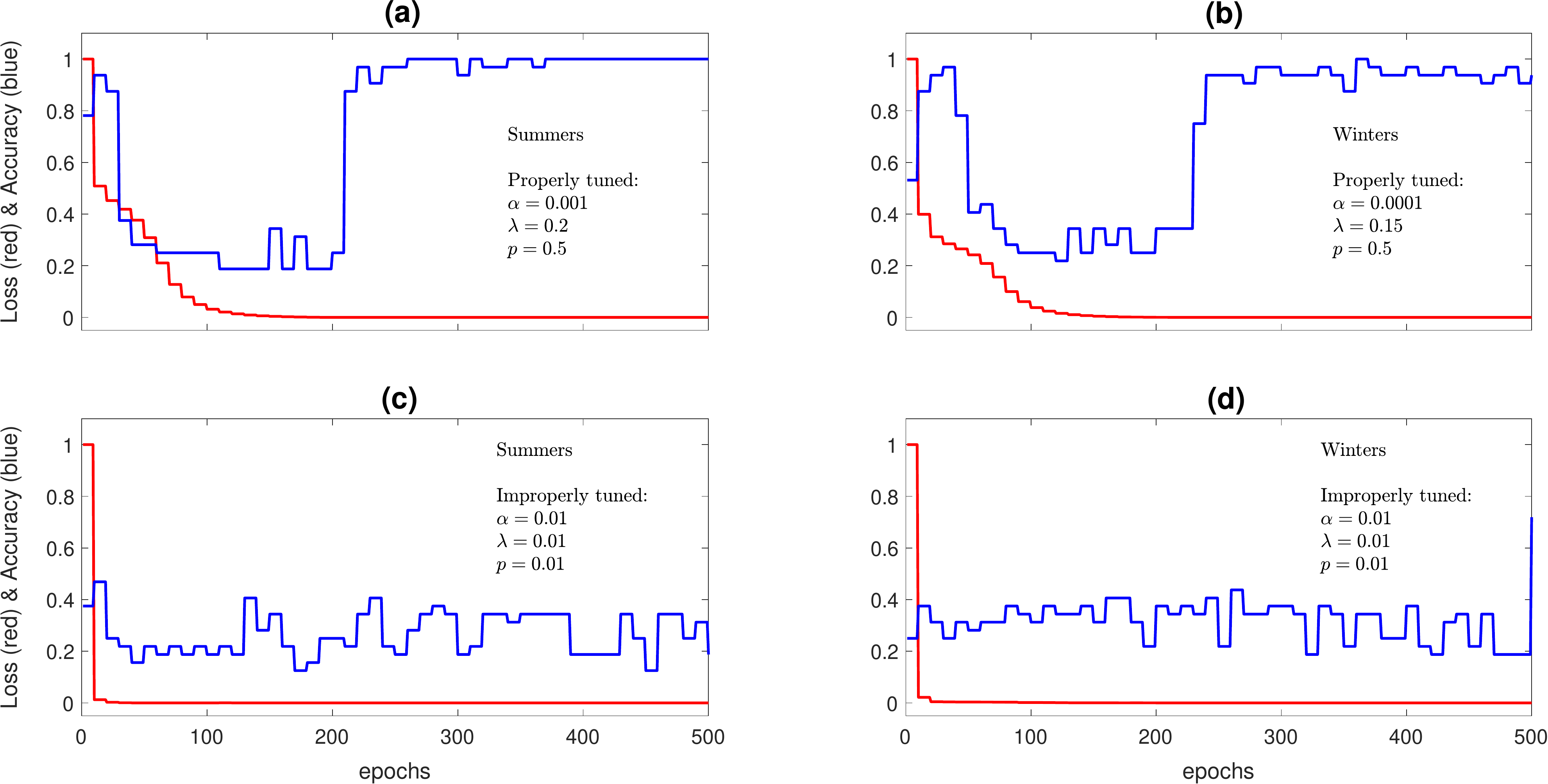}
  \caption{Examples of how loss and accuracy change with epochs during training for CNN4 for properly tuned and improperly tuned CNNs. Loss is measured as cross entropy ($CE$) normalized by its maximum value while the training accuracy is measured by the number of training samples correctly identified at the end of each epoch. Hyperparameters $\alpha$, $\lambda$, and $p$ are, respectively, the initial learning rate, regularization constant, and dropout probability. ({\bf a})  $\alpha=0.001$, $\lambda=0.2$ and $p=0.5$ for summers with the test accuracy of ${93.3\%}$. ({\bf b})  $\alpha=0.001$, $\lambda=0.15$ and $p=0.5$ for winters with the test accuracy of ${93.8\%}$. ({\bf c})  $\alpha=0.01$, $\lambda=0.01$ and $p=0.01$ for summers with the test accuracy of ${25\%}$. ({\bf d}) $\alpha=0.01$, $\lambda=0.01$ and $p=0.01$ for winters with the test accuracy of ${60\%}$). Several kernel sizes were tried and it was found that $5\times 5$ kernel size gives the best validation accuracy and consequently the best test accuracy.}
  \label{fig:lossvacc}
\end{figure}

In the training phase, the images and their labels, in randomly shuffled batches of size $32$, are inputted into the CNN and hyperparameters $\alpha$, $\lambda$, and $p$ are varied until small loss and high accuracy are achieved. Figure~\ref{fig:lossvacc} shows examples of how loss and accuracy vary with epochs for properly and improperly tuned CNNs. Note that only an initial value of $\alpha$ is specified, which is then optimized using the ADAM algorithm. Once the CNN is properly tuned, the $4$ validation sets are used to check the accuracy of CNN in re-identifying the cluster indices. If the accuracy is not high, $\lambda$ and $p$ are varied manually and training/validation is repeated until they both have similarly high accuracy. We found the best test accuracy with the hyperparameters shown in Figure~\ref{fig:lossvacc}(a)-(b). Furthermore, we explored the effect of other hyperparameters such as the number of convolutional layers (from $2$  to $8$) and the kernel sizes (in the range of $5 \times 5$ to $11 \times 11 $) in the convolutional layers on the performance of CNN for this dataset. We found that a network with more than $4$ convolutional layers overfits on $12000$ samples thus producing test accuracy lower than what is reported for CNN4 in Tables~\ref{table:summer} and \ref{table:winter}.  Again, the best test accuracy is found with the architecture shown in Figure~\ref{fig:arch} and described above.      

In the testing phase, the best trained CNN is applied on the $5$ datasets of TS once. The mean and standard deviation of the computed accuracy among these $5$ datasets are reported in Tables~\ref{table:summer} and \ref{table:winter}.  

For the cases with $N=500$ to $8000$, conducted to study the effect of the size of the training set $N$ on the performance of CNN, $N/4$ labeled images from each of the four clusters is selected randomly and used to train the CNN while testing is done on $N/8$ (to the nearest integer) images from each class. 

\subsubsection*{Alternative approach: Applying CNN on data matrix rather than images}
While CNNs are often used on images, they can be used directly to find features in the data matrices as well. For example, we can get the same accuracy as the CNN applied on images with CNN applied on a data matrix of labeled patterns. In such a data matrix $X$, each column contains the full Z500 over $97 \times 66$ grid points for each day. The CNN is applied to $X$, although the best results are obtained with a CNN whose architecture is slightly different from the one applied to images. In this case, the four convolutional layers have $8$, $8$, $16$ and $32$ filters while the fully connected layer has $200$ neurons.
 

\bibliography{main}

\section*{Acknowledgements}
This work was partially supported by NASA grant 80NSSC17K0266 and NSF grant AGS-1552385. Computational resources on the Stampede2 and Bridge GPU clusters and Azure cloud-computing system were provided by the XSEDE allocation ATM170020 and a grant from Microsoft AI for Earth, respectively. A.C. thanks the Rice University Ken Kennedy Institute for a BP HPC Graduate Fellowship. We are grateful to Ashkan Borna, Rohan Mukherjee, Ebrahim Nabizadeh, and Shashank Sonkar for insightful discussions. 



\end{document}